# Active Electrically Small Antenna Based on Superconducting Quantum Array

Victor K. Kornev, Igor I. Soloviev, Alexey V. Sharafiev, Nikolay V. Klenov, and Oleg A. Mukhanov

*Abstract*—We introduce Superconductive Quantum Arrays (SQAs) and propose to use these structures as active Electrically Small Antennas (ESA). Several prototypes of the active ESA were fabricated using Nb process with critical current density 4.5 kA/cm$^2$ and experimentally evaluated. The magnetic field to voltage transfer function linearity up to 70 dB was measured, and transfer factor $dV/dB$ up to 6.5 mV/µT was observed for the ESA prototype containing 560 cells.

*Index Terms*—Active electrically small antenna, arrays, SQUID, Josephson junctions, dynamic range, linearity.

## I. INTRODUCTION

SUPERCONDUCTRING Quantum Arrays are homogeneous arrays of identical cells based on SQUID-type circuits capable of providing a highly linear voltage response to an external magnetic field. Such an array is characterized by an independent operation of these cells which collectively forming an output signal. Two types of the basic cells were proposed, designed, and experimentally evaluated. The first cell is the recently introduced bi-SQUID [1], [2]. The second one is the cell consisting of two parallel Josephson-junction arrays which are differentially connected and oppositely biased by a magnetic flux δΦ [3]-[6]. To get an additional improvement in linearity of the cell voltage response, one can use parallel SQIFs (Superconducting Quantum Interference Filters [7]) instead of a regular array [6]. It is worth noting that Superconducting Quantum Array as a whole is not a SQIF [7]-[10]. Fig. 1 presents a conceptual sketch of a two-dimensional (2D) Superconducting Quantum Array (SQA) and two 3D-configured multichip SQAs.

The SQAs can be used as high efficiency active Electrically Small Antennas (ESAs). Such active ESAs can combine all advantages of passive normal-metal and superconductor electrically small antennas [11]-[19] with high output response and adjustable output impedance. Dynamic range of the active ESA increases with total number of cells in SQA while

Manuscript received October 9, 2012. This work was supported by ONR Program 6.1 under grant RUP1-1493-05-MO via CRDF GAP and in part by grant of Russian Ministry of Science GK 02.740.11.0795 and grant on scientific school PGSS 2456.2012.2.

V. K. Kornev, F. V. Sharafiev, N. V. Klenov are with Physics Department, Moscow State University, Moscow, Russia. (phone: 7-495-939-4351, fax: 7-495-939-3000, e-mail: kornev@phys.msu.ru).
I. I. Soloviev is with Nuclear Physics Institute, Moscow State University, Moscow, Russia (e-mail: isol@phys.msu.ru).
O. A. Mukhanov is with the HYPRES, Elmsford, NY 10523, USA (e-mail: mukhanov@hypres.com).

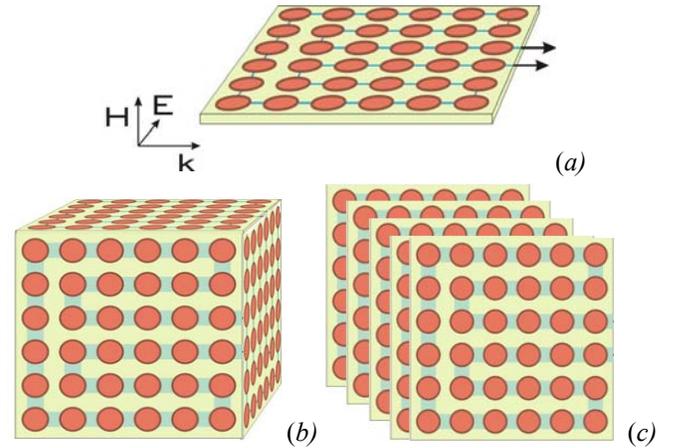

Fig. 1. Conceptual design of (a) a 2D Superconducting Quantum Array (SQA); (b) a 3-chip SQA with orthogonal orientation of SQA array chips; (c) a multi-chip SQA. Such arrays are characterized by an independent operation of all the cells forming a collectively output signal.

linearity of the antenna characteristics is determined by the linearity of the individual magnetic field-to-voltage responses of the cells. For a serial connection of the cells, the output voltage signal $V_{out}$ is proportional to the number $N$ of the cells, while the root-mean-square value of the thermal noise voltage $V_F$ across the array increases as $\sqrt{N}$. For a parallel connection of the cells, output signal $V_{out}$ remains same, and the root-mean-square value of the thermal noise voltage $V_F$ decreases as $\sqrt{N}$. In both cases dynamic range $DR = V_{out}/V_F$ increases as $\sqrt{N}$.

In this paper, we report the results of experimental evaluation the active ESA prototypes fabricated using Nb process with critical current density 4.5 kA/cm$^2$ [20]-[22]. Two possible approaches to the active ESA design were realized. The first one is based on the integration of 1D or 2D SQA with a superconducting flux transformer [6] and the second one is a transformer-free design based on implementation of 2D SQA with the normal-metal cell-to-cell connections.

## II. ACTIVE ESA WITH SUPERCONDUCTING FLUX TRANSFORMER

In case of the transformer-type active ESA, a cell-to-cell connection can be done using both normal metal and superconductor. The array can have a common transformer loop or each cell of the array can be provided with an individual flux transformer.

Fig. 2 shows schematic and microphotograph of the active

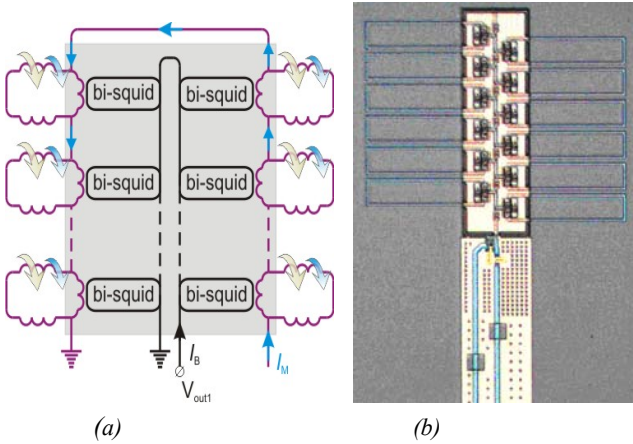

*(a)* *(b)*

Fig. 2. (a) Schematic and (b) microphotograph of the active ESA based on serial array of 12 bi-SQUIDs with individual flux transformer loops of size 0.2 mm x 0.05 mm [1], [6].

ESA based on a serial array of 12 bi-SQUID cells each equipped with an individual 0.2 mm x 0.05 mm flux transformer loop. The gray-filled area in the schematic corresponds to the circuit section with a superconducting ground plane. The array is dc biased by $I_b \approx 2I_C$, where $I_C \approx 125$ µA is Josephson junction critical current, and flux bias current $I_M$ is to set the operation point at the centre of the linear region of the antenna voltage response. An external magnetic field was applied to the antenna prototype by an on-chip one-turn coil circumventing the chip perimeter. Transfer factor dV/dB of the antenna was measured as 50 µV/µT.

Fig. 3(a),(b) shows two microphotographs of a 3.3 x 3.3 mm² central areas of the 5 x 5 mm² chips with the active ESA prototypes. These circuits are designed as two differentially connected serial arrays of 80 ten-junction parallel SQIFs equipped with two superconducting transformer loops and one common transformer loop. Fig. 3(c) shows a schematic for the circuit with a common transformer loop. Inset shows a microphotograph of the array section containing four SQIFs. Each serial array of the SQIFs is biased by current $I_B \approx 10 \cdot I_C$, where Josephson-junction critical current $I_C \approx 125$ µA. An opposite dc magnetic flux biasing of the arrays is achieved by the application of flux biasing "magnetic" current $I_M$ to the opposite points of the transformer loop. This current divides in parallel into the transformer loop shoulders coupled to the arrays. An external magnetic field is applied to the antenna prototypes using an on-chip coil. Transfer factor dV/dB of the antenna prototypes was measured as high as 500 µV/µT and 750 µV/µT, correspondingly.

Fig. 4 shows a set of the voltage responses to the applied magnetic field for the one-loop antenna prototype with the magnetic flux bias current $I_M$. Inset shows a set of individual voltage responses of the arrays of SQIFs with increase in bias current $I_b$ starting from $I_b = 10 \cdot I_c$. To evaluate the linearity of the antenna prototype transfer function, a standard two-tone analysis was performed. This technique is well known and used as for the linearity measurements and analysis of mixing characteristics (e.g., see [23]). Fig. 5 shows a spectrum of the output signal observed at signal frequency 300 kHz. We obtained a 70 dB linearity within ~30% to ~80% of the linear region of differential voltage response.

### III. TRANSFORMER-FREE ACTIVE ESA

A 2D SQA with the normal-metal cell-to-cell connection can be directly used as a high-efficiency active ESA. A homogeneity of the structure ensures a uniform distribution of the B-field component of incident electromagnetic wave (except the side rows which could be excluded for this reason). The serial connection of the cells, each providing linear response, results in a linear voltage output with the N times increase in transfer factor and the $N^{1/2}$ times increase in dynamic range, where N is a total number of cells.

Fig. 6 explains the mechanism of the magnetic flux

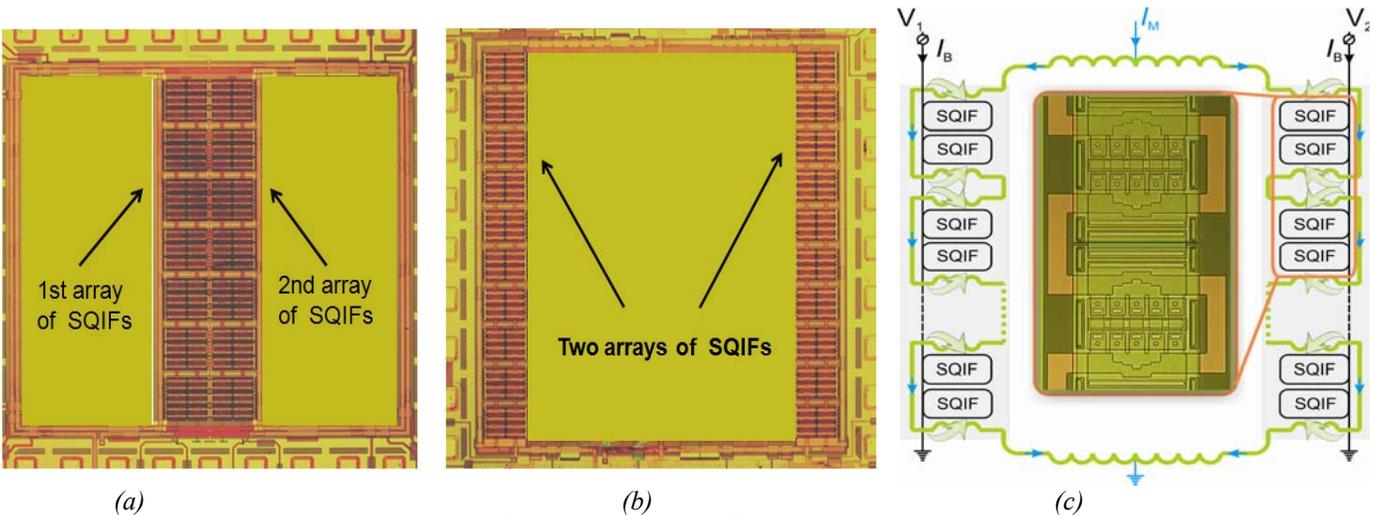

*(a)* *(b)* *(c)*

Fig. 3. Microphotographs of the central area 3.3 x 3.3 mm² of chip of size 5 x 5 mm² with active ESA prototypes based on two differentially connected serial arrays of 80 ten-junction parallel SQIFs with (*a*) two superconducting transformer loops and (*b*) a single common transformer loop as well as (*c*) a schematic for the chip design depicted in (*b*) [6]. Inset shows a microphotograph of the array section containing 4 SQIFs. $I_B$ is biasing current; $I_M$ is "magnetic" current providing opposite dc magnetic flux biasing of the antenna array shoulders.

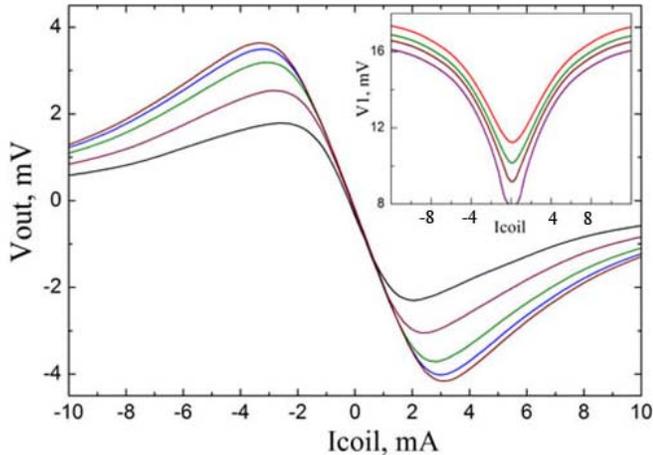

Fig. 4. Set of voltage responses [6] of the shown in Fig. 3(b) antenna at dc current biasing of the serial arrays of ten-junction SQIFs $I_B = 10 \cdot I_C$, where $I_C = 125$ μA is Josephson junction critical current. Amplitude of the responses increases with "magnetic" current $I_M$ providing opposite dc magnetic flux biasing of the differentially connected arrays of SQIFs. $I_{coil}$ is the coil current producing external magnetic field. Inset shows a set of individual voltage responses of the arrays with bias current $I_B$, starting from $I_B = 10 \cdot I_C$ (lower curve).

application to the SQIF-based cell, when magnetic vector $H$ of the incident electromagnetic wave is perpendicular to the cell plane. Meissner currents flow on borders of the M1 and M2 superconductor layers and connect on the interior surfaces in the layer intersection. These two opposite currents (see a cross section view) induce a magnetic flux to a SQIF-cell. In fact, magnetic field $B$ in the gap between the superconductor films meets equation $rot(B) = \mu_0 j$, where $j$ is density of the current. Therefore, the applied to every section of the SQIF magnetic flux equals to $B \cdot d \cdot dx$, where $d$ is thickness of isolator film, and $dx$ is a distance between Josephson junctions.

Fig. 7 shows a microphotograph of the fabricated and experimentally evaluated transformer-free antenna prototype containing 1120 single-SQIF cells (i.e., 560 differential cells) and occupying a 3.3 x 3.3 mm$^2$ area on a standard 5 x 5 mm$^2$ chip. Each SQIF consists of 12 identical Josephson junctions (Josephson junction critical current $I_c = 125$ μA) connected in parallel. The cells are sectioned into two serial arrays of 560 single-SQIF cells with normal-metal cell-to-cell coupling. The arrays are differentially connected to the antenna output. To realize an opposite dc magnetic flux biasing of the arrays, each array is equipped with a control line with normal-metal sections between the SQIF-cells. An external multi-turn coil was used to apply magnetic field to the antenna prototype.

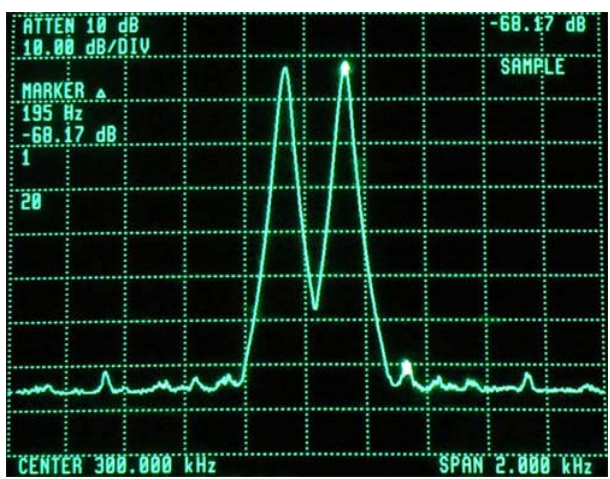

Fig. 5. Result of two-tone analysis of the antenna prototype voltage response at signal frequency 300 kHz [6]. The analysis gives linearity of approx. 70 dB within 30% to 80% of the voltage response swing in dependence on magnetic frustration of the antenna arrays.

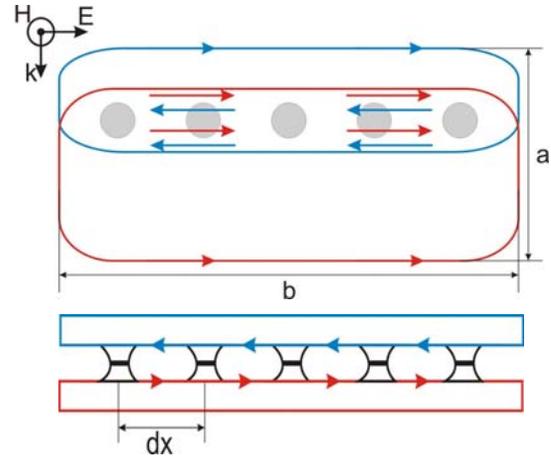

Fig. 6. Schematic explanation of a mechanism of magnetic flux application to the SQIF-based cell when magnetic vector $H$ of the incident electromagnetic wave is perpendicular to the cell plane. Meissner currents flow on borders of the M1 and M2 superconductor layers and connect on the interior surfaces in the layer intersection.

Typical magnetic field to voltage transfer function of the transformer-free antenna prototype and a set of the curves with magnetic opposite biases of the antenna array shoulders are presented in Fig. 8. Maximum voltage response amplitude approaches ~80 mV, and transfer factor was measured as high as ~6.5 mV/μT.

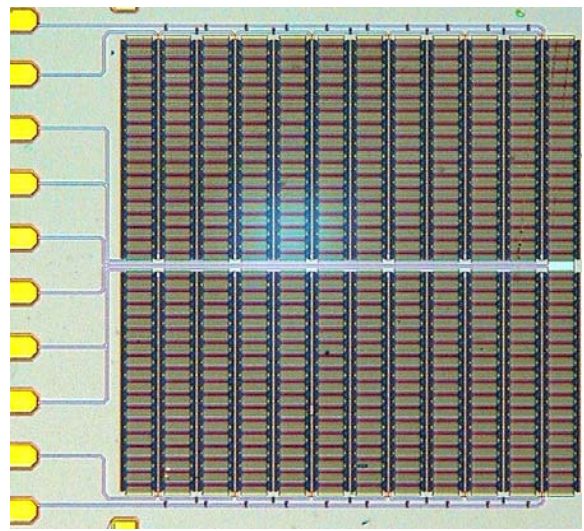

Fig. 7. Microphotograph of the transformer-free antenna prototype containing 1120 single-SQIF cells (i.e., 560 differential cells) and occupying 3.3 x 3.3 mm$^2$ on a 5 x 5 mm$^2$ chip.

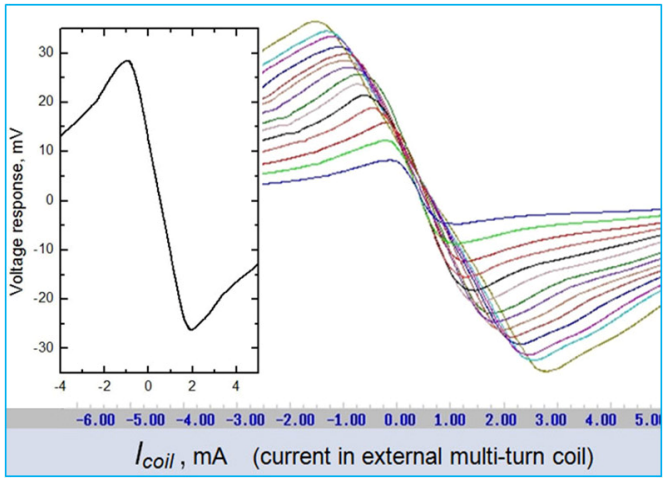

Fig. 8. Typical magnetic field to voltage transfer function curve (a middle curve of the set shown) for the transformer-free antenna prototype and a set of the curves at dc current biasing of the serial arrays of twelve-junction SQIFs $I_B \approx 12 \cdot I_C$, where $I_C = 125$ µA is Josephson junction critical current. Amplitude of the responses increases with increase of the opposite dc magnetic flux biasing (magnetic frustration) of the antenna array shoulders. $I_{coil}$ is current via external multi-turn coil producing external magnetic field.

## IV. Conclusions

Several prototypes of active Electrically Small Antennas (ESA) based on Superconducting Quantum Arrays (SQA) were designed, fabricated and experimentally evaluated. Two approaches to the design of the antennas were developed and realized: (i) an integration of a 1D SQA with superconducting flux transformer and (ii) a transformer-free design based on the implementation of a 2D SQA with normal-metal cell-to-cell coupling. Both versions occupy the same area of 3.3 x 3.3 mm$^2$. Our antenna prototypes with the transformer and transformer-free designs showed transfer factors of ~750 µV/µT and ~6.5 mV/µT, correspondingly.

As it is pointed out above, the $B$ to $V$ transfer factor of a serial array of SQIFs is proportional to number $N$ of SQIFs, whereas the root-mean-square value of thermal noise voltage across the array increases as $\sqrt{N}$, and hence dynamic range increases as $\sqrt{N}$. In turn, this means that dynamic range of the active ESA occupying square area $A$ x $A$ increases as $\sqrt{A}$ in the case of a 1D SQA integrated with square transformer loop, and as $A$, i.e., much faster, in case of the transformer-free active ESA based on a 2D SQA.

One can expect that the implementation of a 3D SQA or a set of three 2D SQA with mutually perpendicular orientations (Fig. 1) as an active ESA should allow the improvement of all characteristics of the antenna including its transfer factor, dynamic range, output impedance and directional pattern.

Such active high-sensitive broadband superconductor antenna is in demand for the superconductor Digital-RF receiver systems with direct digitization [24]-[30]. The present-day superconductor ADCs demonstrate outstanding linearity and dynamic range [31]-[35]. However, the inferior linearity and dynamic range of antennas and low noise amplifier compared to those of the ADCs constrain the overall system performance. Development of the high-efficiency active ESA based on implementation of SQAs will help to overcome these limitations and improve the overall system performance.

## Acknowledgment

Authors thank D. Van Vechten for the attention to this work, Hypres fabrication team of D. Yohannes, J. Vivalda, R. Hunt, D. Donnelly, D. Amparo for manufacturing the integrated circuits, A. Kirichenko, T. Filippov and D. Kirichenko for help and valuable discussions.